\documentstyle[prd,eqsecnum,aps,epsf]{revtex}

\newcommand{\singlefig}[2]{
\begin{center}
\begin{minipage}{#1}
\epsfxsize=#1
\epsffile{#2}
\end{minipage}
\end{center}}
\newenvironment{figcaption}[2]{
 \vspace{0.3cm}
 \refstepcounter{figure}
 \label{#1}
 \begin{center}
 \begin{minipage}{#2}
 \begingroup \small FIG. \thefigure: }{
 \endgroup
 \end{minipage}
 \end{center}}

\begin{document}

\newcommand{\beq}{\begin{equation}}
\newcommand{\eeq}{\end{equation}}
\newcommand{\bea}{\begin{eqnarray}}
\newcommand{\eea}{\end{eqnarray}}
\newcommand{\p}{{\varphi}}
\newcommand{\pd}{\dot{\varphi}}
\newcommand{\pdd}{\ddot{\varphi}}
\newcommand{\ld}{\dot{\lambda}}
\newcommand{\ldd}{\ddot{\lambda}}
\newcommand{\nd}{\dot{\nu}}
\newcommand{\ndd}{\ddot{\nu}}
\newcommand{\tn}{\tilde{\nu}}

\preprint{BROWN-HET-....}

\title{Isotropization in Brane Gas Cosmology}

\author{Scott Watson and Robert H. Brandenberger
\footnote{email: watson@het.brown.edu, rhb@het.brown.edu}}

\address{Department of Physics, Brown University, Providence, RI 02912, USA}

\date{\today}

\maketitle

\begin{abstract}
Brane Gas Cosmology (BGC) is an approach to unifying string theory
and cosmology in which matter is described by a gas of strings and
branes in a dilaton gravity background. The Universe is assumed
to start out with all spatial dimensions compact and small. It
has previously been shown  that in this context,
in the approximation of neglecting inhomogeneities and anisotropies,
there is a dynamical
mechanism which allows only three spatial dimensions to become
large. However, previous studies do not lead to any conclusions
concerning the isotropy or anisotropy of these three large
spatial dimensions. Here, we generalize the equations of BGC
to the anisotropic case, and find that isotropization is a natural
consequence of the dynamics.
\end{abstract}

\pacs{Valid PACS appear here}

\section{Introduction}

The Standard Big Bang (SBB) cosmology has become an extremely successful
model that has been well tested by experiment. However,
the model is incomplete.  The underlying theory of classical
general relativity and the description of matter as an ideal gas breaks down
at the high temperatures of the early Universe, and the solutions
of the theory in fact have an initial singularity. Moreover, SBB does
not address many important cosmological questions such as the observed
homogeneity, spatial flatness, and the origin of structure in the
Universe.  Cosmological inflation (see e.g. \cite{Linde:nc,Liddle:cg} for
textbook treatises and \cite{Watson:2000hb,Brandenberger:1999sw} for
shorter reviews) builds on SBB cosmology providing a solution to some of these
issues, but it
(at least in the context of scalar field-driven inflation)
suffers from the same initial singularity problem \cite{Borde:xh}
and other conceptual problems \cite{Brandenberger:1999sw}, which
indicate that inflation cannot be the complete story of early
Universe cosmology.

In recent years, many models motivated by string theory and M-theory have
emerged as possible solutions to the
outstanding problems of early Universe and inflationary cosmology
(see e.g. \cite{Easson:2000mj,Brandenberger:2001ph} for recent but
incomplete reviews). Beginning with the work on Pre-Big-Bang cosmology
\cite{Veneziano:1991ek,Gasperini:1992em} it was realized that a dynamical
dilaton should play an important role in the very early Universe. More
recently, models have become prominent in which  our Universe consists
of a 3-brane embedded in a higher dimensional bulk space, with the
standard model constrained to live on the brane
\cite{Arkani-Hamed:1998rs,Antoniadis:1998ig,Randall:1999vf,Randall:1999ee,Khoury:2001wf,Kallosh:2001ai,Steinhardt:2001vw}.
Although these models can resolve a number of issues, such as the hierarchy
problem, they introduce several other difficulties in the process.
For example, large extra dimensions should be explained by
classical general relativity, and it has been shown this results in problems
stabilizing the brane \cite{Carroll:2001ih}.
More importantly, in most of these models the six ``extra'' spatial
dimensions are taken to be
compactified, {\em a priori}, with no explanation provided for how this could
come about dynamically.  Although
it appears to be an important concern for the naturalness of
these models, this issue is rarely discussed in the literature.

An alternative approach to string/M-theory cosmology is the string gas
or BV scenario. This model began with works
\cite{Brandenberger:1989aj,Tseytlin:1992xk}
in which the effects of string gases on the cosmological evolution of the low
energy effective string theory background geometry including the dilaton
were explored.  The most important result to emerge from
these works is a dynamical mechanism, tied to the existence of
string winding modes, which yields a nonsingular cosmology and
may explain why at most three spatial dimensions can become large
if the initial state is chosen to correspond to a Universe which is
small in all spatial directions.

This work has been generalized to include the cosmological effects
of p-brane gases and leads to
the current model of Brane Gas Cosmology (BGC) \cite{Alexander:2000xv}.
In BGC, the Universe starts analogous to the SBB picture, i.e. hot, dense,
and with all fundamental degrees of freedom in thermal equilibrium.
The Universe is
assumed to be toroidal in all nine spatial dimensions and filled
with a p-brane gas. The assumption of toroidal
geometry of the background space leads to the existence of
string winding modes, since the background space  admits cycles
on which  branes of the relevant dimensionalities (in particular
one branes) can wrap. This wrapping is associated with a
winding energy which - in the context of dilaton gravity - acts as a
confining potential for the scale factor preventing
further expansion of the spatial dimensions.  Also associated with the
brane are oscillatory modes described by scalar fields and momentum
modes which correspond to the center of mass motion of the brane.
The momentum modes are related by T-duality to the winding modes
and this duality results in the non-singular behavior of the model.
In order for dimensions to decompactify, p-brane winding
modes must annihilate with anti-winding modes and it is argued that this
only occurs in a maximum of $2p+1$ dimensions \cite{Alexander:2000xv}.
Since strings $(p=1)$ are the lowest dimensional objects that admit winding
modes, since they are the lightest of all winding modes and hence fall
out of equilibrium later than other winding modes,
it follows that the number of large space-time
dimensions can be at most (3+1).

In the context of the background equations of dilaton gravity, the
winding modes yield a confining potential for the scale factor which also
gives rise to a period of cosmological loitering (expansion rate near zero)
for the three large spatial dimensions. This is due to the time needed for
winding modes to annihilate
and produce closed strings or loops \cite{Brandenberger:2001kj}.
This is of great interest since loitering can explain the
horizon and relic problems of standard
cosmology without resorting to inflation
\footnote{However, to obtain a solution of the flatness and entropy
problems, a phase of inflation following the decoupling of the three
large spatial dimensions may be required}.
It was also shown in \cite{Brandenberger:2001kj}
that by considering loop production at late times BGC naturally
evolves into the SBB, with a $3+1$ dimensional, radiation dominated Universe.

There are important issues that remain to be addressed
within BGC.  The fact that toroidal geometry was assumed for the background
space was used for the existence and topological stability of winding modes.
However, K3 or Calabi-Yau manifolds
are more realistic choices for backgrounds within string theory and they
do not admit 1-cycles (necessary to have topologically stable
winding modes). Promising results have recently appeared which
indicate that the conclusions of BGC extend to a much wider class of
spatial background, including
backgrounds which are K3 or Calabi-Yau manifolds
\cite{Easson:2001re,Easther:2002mi}.

Perhaps the main issue to be addressed in BGC is that of spatial
inhomogeneities.  That is, we would naturally expect fluxes and
p-brane sources to lead to the possibility of catastrophic
instabilities of spatial fluctuation modes. Although we do not address this
issue here, we plan to study the role of inhomogeneities in followup work.

Other important issues for BGC are the questions of stabilization
of the six small extra dimensions and isotropization of the three
dimensions that
grow large.  Although these topics may seem to be unrelated, they
both can be addressed by generalizing the BGC scenario to the
anisotropic case.  This paper will concentrate on the
isotropization of the three large dimensions, but the generalization of
BGC to the anisotropic case achieved in this paper
will be valuable to address the issue of stabilization in later work.

\section{Brane Gas Cosmology}

We begin with a brief review of BGC, for more details the reader
is referred to \cite{Alexander:2000xv}.
Consider compactification of 11-dimensional
M-theory on $S^{1}$, which yields type II-A string theory in
10-dimensions.  The fundamental degrees of freedom in the theory
are 0-branes, strings, 2-branes, 5-branes, 6-branes,
and 8-branes. The low energy effective action of the theory is
that of supersymmetrized dilaton gravity,

\beq \label{Sbulk}
S_{bulk} = \frac{1}{2 \kappa^2}\int d^{10}x \sqrt{-G} e^{-2 \phi}
\bigl[ R + 4 G^{\mu \nu} \nabla_\mu \phi \nabla_\nu \phi
- \frac{1}{12} H_{\mu \nu \alpha} H^{\mu \nu \alpha} \bigr]
\,,
\eeq
where $G$ is the determinant of the background metric $G_{\mu \nu}$,
$\phi$ is the dilaton given by the radius of the $S^{1}$
compactification, $H$ denotes the field strength corresponding to
the bulk antisymmetric tensor field $B_{\mu \nu}$, and $\kappa$ is
determined by the ten-dimensional Newton constant.

Fluctuations of each of the $p$-branes are described by the Dirac-Born-Infeld
(DBI) action~\cite{Polchinski:1996na}
and are coupled to the ten-dimensional action via delta function sources.
The DBI action is
\begin{equation} \label{Sbrane}
S_p \, = \, T_p \int d^{p + 1}
\zeta e^{- \phi} \sqrt{- det(g_{mn} + b_{mn} + 2 \pi \alpha' F_{mn})}
\,,
\end{equation}
where $T_p$ is the tension of the brane, $g_{mn}$ is the induced metric on
the brane, $b_{mn}$ is the induced antisymmetric tensor field, and $F_{mn}$
is the field strength tensor of gauge fields $A_m$ living on the brane.  The
constant $\alpha' \sim l_{st}^2$ is given by the string length scale
$l_{st}$.

In this paper we will concentrate on the role of fundamental
strings ($p=1$), since strings
play the critical role in the dynamics \cite{Alexander:2000xv}.
We will also ignore the
effects of fluxes and string oscillatory modes since we are
working in the low energy regime.
Our basic approach is to consider the string gas as a matter source for
the dilaton-gravity background. The string winding modes play a key
role. The main point is that we want
to find the first departures from the standard FRW picture resulting
from considering {\em stringy} effects.  This should result from the
winding modes of the strings
and the presence of the dilaton in the equations of motion.

In this paper, we generalize the equations of BGC to the anisotropic
case. Thus, we take the metric to be of the form
\begin{equation} \label{metric}
ds^2 \, = \, dt^2 - \sum_{i = 1}^{D} e^{2\lambda_i} (dx_i)^2 \, ,
\end{equation}
where the label $i$ runs over the $D$ spatial indices, $x_i$ are the
comoving spatial coordinates, $t$ is physical time, and the scale
factor in the i'th direction is $log(\lambda_i)$. We again stress that
inhomogenities are of vital importance, however we will leave their
investigation to future work and take the string gas to be homogeneous.

By varying the total action we obtain the following equations,
\begin{eqnarray}\label{eom}
-\sum_{i=1}^{D} {\dot \lambda}_{i}+\pd^{2}=e^{\p}E\\
-\sum_{i=1}^{D} {\dot \lambda}_{i}+\ddot{\p}=\frac{1}{2} e^{\p} E\\
\ldd_{i}-\pd\ld_{i}=\frac{1}{2}e^{\p}P_{i}\label{eq3},
\end{eqnarray}
where $E$ is the total energy of the strings and $P_{i}$ is the pressure in
the $ith$ direction, and we have introduced a ``shifted'' dilaton field
\begin{equation} \label{shift}
\varphi \, = \, 2 \phi - \sum_{i = 1}^{D} \lambda_i \, .
\end{equation}

We can see immediately from the first equation that $\varphi$ can not
change sign.  To insure that the low energy approximation remains
valid we choose $\p<0$ and $ {\dot \varphi}>0$. This causes the dilaton to have a
damping effect in the equations of motion
\footnote{Alternatively we could consider $\p>0$ and one might
anticipate that this could lead to inflation.  However this is a false
conclusion, since the presence of winding
modes act as a confining potential which can be seen from the negative
pressure term in equation (\ref{eq3}).}.
Another important observation is that these equations reduce to
the standard FRW equations for fixed dilaton as was discussed in
\cite{Tseytlin:1992xk}.  This is comforting because at late times it
is expected that the dilaton will gain a mass, perhaps associated
with supersymmetry breaking and as a result we naturally
regain the usual SBB cosmology. The string winding states will appear
at late times as solitons \cite{Alexander:2000xv}.

\subsection{Winding Modes}

The energy and pressure terms follow from varying the brane
action (\ref{Sbrane}).
The energy associated with the winding modes in the $ith$
direction is given by (taking the winding strings to be
straight for simplicity)
\begin{equation}
E^{w}_{i}=\mu N_{i}(t) a_{i}(t)=\mu N_{i}(t) e^{\lambda_{i}},
\end{equation}
where $N_{i}$ is the number of winding modes in the {\em i}th direction
and $\mu$ is the mass per unit length of the string multiplied by the
initial spatial dimension. It follows that the total energy is given by,
\begin{equation}\label{wenergy}
E^{w}_{T}=\sum_{i=1}^{D} \mu  N_{i}(t) e^{\lambda_{i}}.
\end{equation}
In the same approximation that the winding strings are
straight, the corresponding pressure terms are given
by \cite{Alexander:2000xv},
\begin{eqnarray} \label{wpressure}
P^{w}_{i} &=& -\mu N_{i}(t) e^{\lambda_{i}} \, \\
P^{w}_{j} &=& 0 \,\,\, (j \neq i) \, .
\end{eqnarray}
It follows then by inserting (\ref{wpressure}) into (\ref{eq3}) that
winding strings give rise to a confining potential for the scale
factor, and hence prevent expansion. In order for a spatial dimension
to be able to become large, the winding modes in that direction must
be able to annihilate.

\subsection{Annihilation and Loop Production}

As argued earlier, the fact that strings are likely to annihilate
in a maximum of three space dimensions leads to three of the initial
nine spatial dimensions growing large.  As the background
continues to expand the three dimensional space will be filled
with loops resulting from the annihilation of the winding states.
The corresponding energy loss in winding states and the energy
and pressure of the creation of loops must be taken into account
when considering the dynamics.  The strings in the expanding space
will become of macroscopic size and the required pressure and energy
terms will be analogous to that of a cosmic string network
\cite{Vilenkin:1994,Brandenberger:1993by}.

We know that the energy in winding modes is given by
(\ref{wenergy}).  Considering the time rate of change of this
energy we find,
\beq \label{changeEw}
{\dot E}^{w}_{T}=\sum_{i=1}^{D} \mu {\dot N}_{i}(t) e^{\lambda_{i}}
+\sum_{i=1}^{D} \mu { N}_{i}(t)
e^{\lambda_{i}}\ld_{i}.
\eeq
The first term on the right hand side of
this equation corresponds to energy loss into string loops or
radiation, the second term corresponds to the gain in total
energy of the strings due to the stretching by the expansion
of space.

The energy in loops can be written as \cite{Brandenberger:2001kj}
\beq
{ E}^{loops}=g(t)V_{0},
\eeq
where $g(t)$ is the energy density in string loops (plus radiation)
per initial comoving volume (and is a constant if no loop production
or energy loss from winding strings into radiation occurs), and
$V_{0}=\exp(\sum \lambda_{0i})$ is the initial volume, $\lambda_{0i}$
being the initial values of the logarithms of the scale factors.
Since sting loops are likely to be produced relativistically, we
will use the equation of state of radiation to describe them. This
approximation also allows us to treat the string loops and other
radiation together. This approximation could easily be relaxed
without changing our basic conclusions. Thus, we use the equation
of state of relativistic radiation $p=(1/3)\rho$
($\rho$ and $p$ denoting energy density and pressure, respectively)
which implies,
\beq
P^{loops}=\frac{1}{3}g(t)V_{0}.
\eeq

To find how $g(t)$ evolves we equate the loss in winding energy
due to energy transfer (the first term on the right hand side of
\ref{changeEw}) with the change in loop energy,
\beq
-\sum_{i=1}^{D} \mu {\dot N}_{i}(t) e^{\lambda_{i}}={\dot g(t)}
V_{0}.
\eeq
Thus, the loop production is determined by the change in the number
of winding modes, as expected:
\beq \label{geqn}
{\dot g(t)}=-V_{0}^{-1}\sum_{i=1}^{D} \mu {\dot N}_{i}(t) e^{\lambda_{i}}.
\eeq
It remains to determine the rate of winding mode annihilation, but
before doing so let us consider the specific case of a 2+1
anisotropic Universe.

\section{Anisotropic Generalization}

As three of the nine spatial dimensions grow large there is no
{\em a priori} reason to expect that this should happen in an
isotropic manner.  Moreover, since the dimensions expanding correspond
to the annihilation of winding modes due to string intersections, we
might expect this process to occur at differing rates leading to
anisotropic dimensions.
However, we might also expect that as the winding modes annihilate
in one dimension and that dimension begins to expand faster there
are then fewer winding modes left to annihilate.  In this way the
remaining dimensions are given an opportunity to isotropize.  To
explore if this is indeed the case, let us consider the case of a
2+1 anisotropic background, whose metric we write in the form
\begin{equation} \label{metric2}
ds^2 \, = \, dt^2 - e^{2\lambda}(dx^2 + dy^2) - e^{2\nu}dz^2 \, .
\end{equation}
The scale factor corresponding to $\lambda$ is denoted $a(t)$, the
one corresponding to $\nu$ is denoted $b(t)$.

The equations of motion (\ref{eom}) then become,
\begin{eqnarray}
-2\ld^{2}-\nd^{2}+\pd^{2}=e^{\p}E\\
-2\ld^{2}-\nd^{2}+\ddot{\p}=\frac{1}{2} e^{\p} E\\
\ldd-\pd\ld=\frac{1}{2}e^{\p}P_{\lambda}\\
\ndd-\pd\nd=\frac{1}{2}e^{\p}P_{\nu} \, .
\end{eqnarray}

The energy and pressure terms are given by
\bea
E = E^{w}_{T}+E^{loops}
& = & 2 \mu  N(t) e^{\lambda}+\mu  M(t)
e^{\nu}+g(t)e^{2\lambda_{0}+\nu_{0}},\\
P_{\lambda} & = & -\mu N(t) e^{\lambda}+\frac{1}{3}g(t)e^{2\lambda_{0}
+\nu_{0}}, \\
P_{\nu} & = & -\mu M(t)e^{\nu}+\frac{1}{3}g(t)e^{2\lambda_{0}+\nu_{0}},
\eea
where $N(t)$ and $M(t)$ are the number of winding modes in the $\lambda$ and
$\nu$ directions, respectively.

We now consider the effect of loop production as a result of
winding mode annihilation. To simplify the analysis, we will
assume that winding modes in $\lambda$ and $\nu$ directions
form two separate, noninteracting gases. We expect that this
approximation will {\it reduce} the isotropization of winding
modes, and hence that including the interactions we omit
will lead to accelerated isotropization. This issue is being
studied currently \cite{SBW}.

Winding mode annihilation results from
string intersections, which depends
\cite{Vilenkin:1994,Brandenberger:1993by} on the square of the number of
modes, inversely on the cross-sectional area available for the
interaction and is proportional to the Hubble length of the string.
Since causality plays a role in the interaction, we must scale our
dimensions by the Hubble time $t$. Hence,
\beq
{\dot N}(t) = -{\tilde c}_{N} N^{2}(t) \Bigl( \frac{t^{2}}{Area}  \Bigr)
t^{-1} = - {\tilde c}_{N} N^{2}(t) \Bigl( \frac{t^{2}}{a(t)b(t)}  \Bigr)
t^{-1} = -c_{N} t N^{2}(t) e^{-\lambda-\nu} \, ,
\eeq
where ${\tilde c}_N$ is a dimensionless constant, and $c_N$ is this same
constant rescaled by the basic length dimensions related to $a$ and $b$.
Similarly, for $M(t)$ we have
\beq
{\dot M}(t) = -c_{M} t M^{2}(t) e^{-2\lambda}.
\eeq

Using these expressions in (\ref{geqn}) gives us the evolution of loop
production,
\beq
{\dot g}(t)=\mu t e^{-2\lambda_{0}-\nu} \Bigl( 2c_{N}N^{2}(t) e^{-\nu}+c_{M}M^{2}(t)
e^{-2\lambda+\nu}\Bigr).
\eeq
For computational simplicity we define $l(t) \equiv \ld$, $q(t)
\equiv \nd$, and $f(t) \equiv \pd$, which leaves us with the
following set of first order ordinary differential equations,
\bea
-2l^{2}-q^{2}+f^{2}=e^{\p}\Bigl( 2 \mu  N(t) e^{\lambda}+\mu  M(t)
e^{\nu}+g(t)e^{2\lambda_{0}+\nu_{0}}  \Bigr) \label{constraint} \\
{\dot f}=2l^{2}+q^{2}+\frac{1}{2}e^{\p}\Bigl( 2 \mu  N(t) e^{\lambda}+\mu  M(t)
e^{\nu}+g(t)e^{2\lambda_{0}+\nu_{0}}  \Bigr)\\
{\dot l}=l\p+\frac{1}{2}e^{\p}\Bigl(-\mu N(t)e^{\lambda}+\frac{1}{3}g(t)e^{2\lambda_{0}+\nu_{0}}  \Bigr)\\
{\dot q}=q\p+\frac{1}{2}e^{\p}\Bigl(-\mu M(t)e^{\nu}+\frac{1}{3}g(t)e^{2\lambda_{0}+\nu_{0}}  \Bigr)\\
{\dot N}(t)=-c_{N} t N^{2}(t) e^{-\lambda-\nu}\\
{\dot M}(t)=-c_{M} t M^{2}(t) e^{-2\lambda}\\
{\dot g}(t)=\mu t e^{-2\lambda_{0}-\nu} \Bigl( 2c_{N}N^{2}(t) e^{-\nu}+c_{M}M^{2}(t)
e^{-2\lambda+\nu}\Bigr) V_0^{-1}\\
l(t) = \ld \\
q(t) = \nd \\
f(t) = \pd. \label{last}
\eea

The system (\ref{constraint}-\ref{last}) is
over-determined (there are more equations than
unknowns). Therefore we take (\ref{constraint}) as a constraint on the
initial conditions. We find from numerical analysis that the
winding modes $N(t)$ and $M(t)$ approach zero as the system
evolves in time, as expected from winding mode annihilation
(see Fig. \ref{swfig1}).  As a result loop
production continues until all the winding modes are
annihilated at which time $g(t)$ approaches a constant (see Fig.
\ref{swfig2}).  Note from Figure \ref{swfig1} that before
converging to zero, the two winding numbers converge to eachother. This
is a necessary condition for isotropization.
This result seems
reasonable, since the rate of annihilation depends on the number of
winding states present.  Thus, as the number of winding modes
decreases in one direction, the other direction should have an
larger rate of annihilation. This suggests a sort of
equilibration process due to the presence of winding modes.

\begin{figure}
\begin{center}
\singlefig{2.5in}{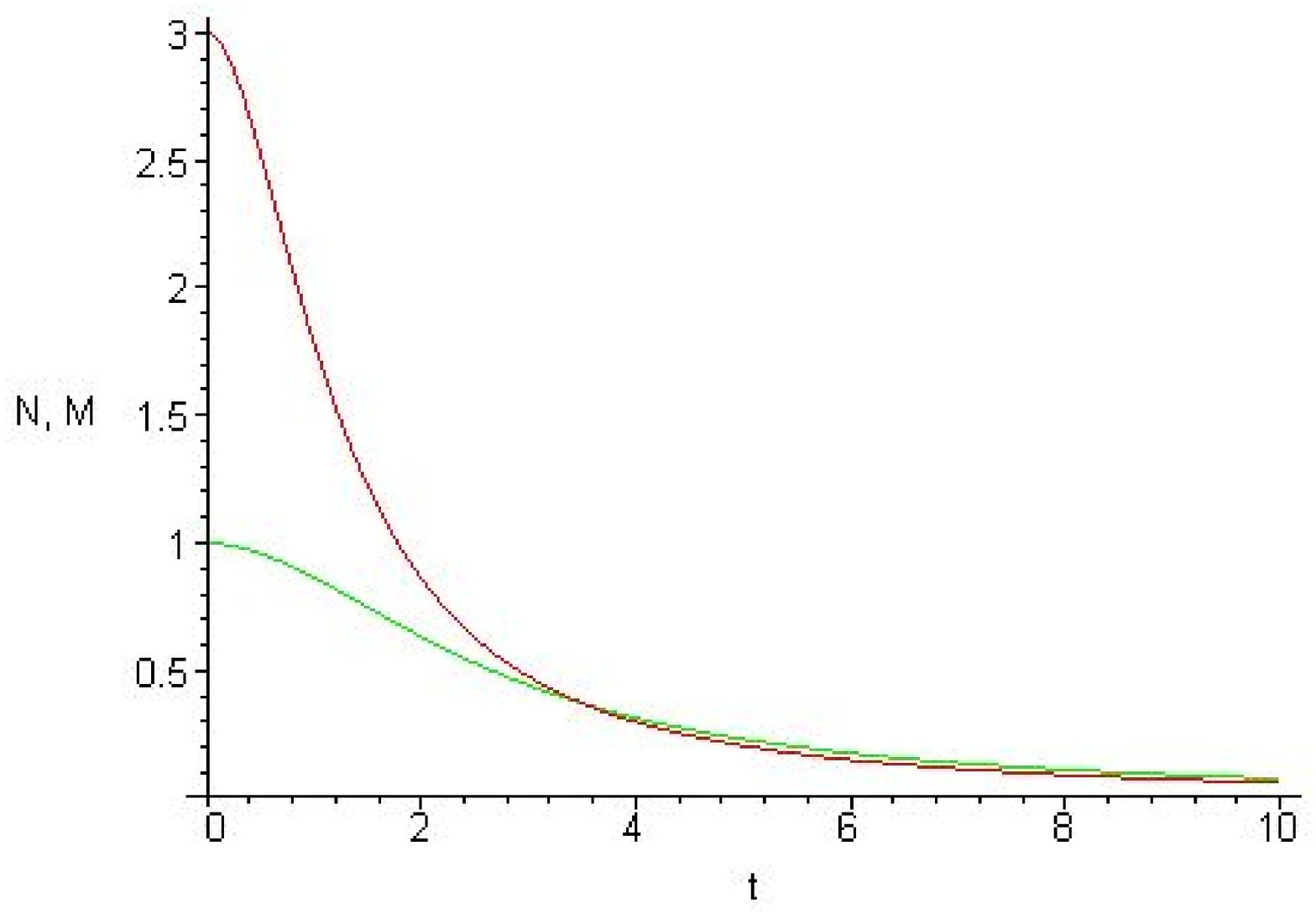}
\begin{figcaption}{swfig1}{2.5in}
A plot of the numbers $N$ (upper curve for small values of $t$)
and $M$ (lower curve) as a
function of time $t$. We see that the number of winding modes approaches
zero as winding annihilation continues.
\end{figcaption}
\end{center}
\end{figure}

\begin{figure}
\begin{center}
\singlefig{2.5in}{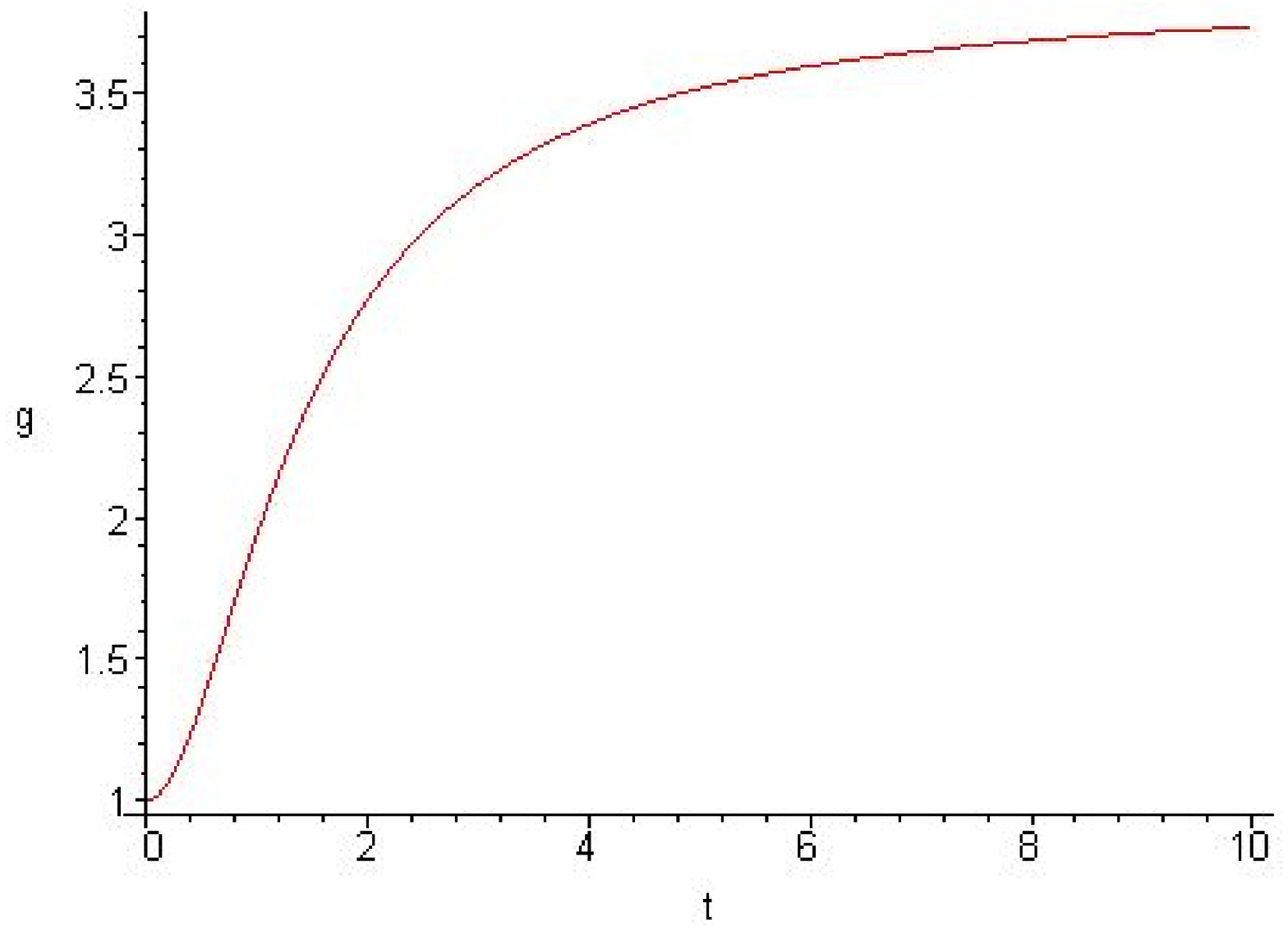}
\begin{figcaption}{swfig2}{2.5in}
A plot of the comoving energy density in loops $g$ as a function of time $t$.
As the winding modes annihilate, loop
production continues until all the winding modes have vanished
and the loop energy becomes constant.
\end{figcaption}
\end{center}
\end{figure}

\begin{figure}
\begin{center}
\singlefig{2.5in}{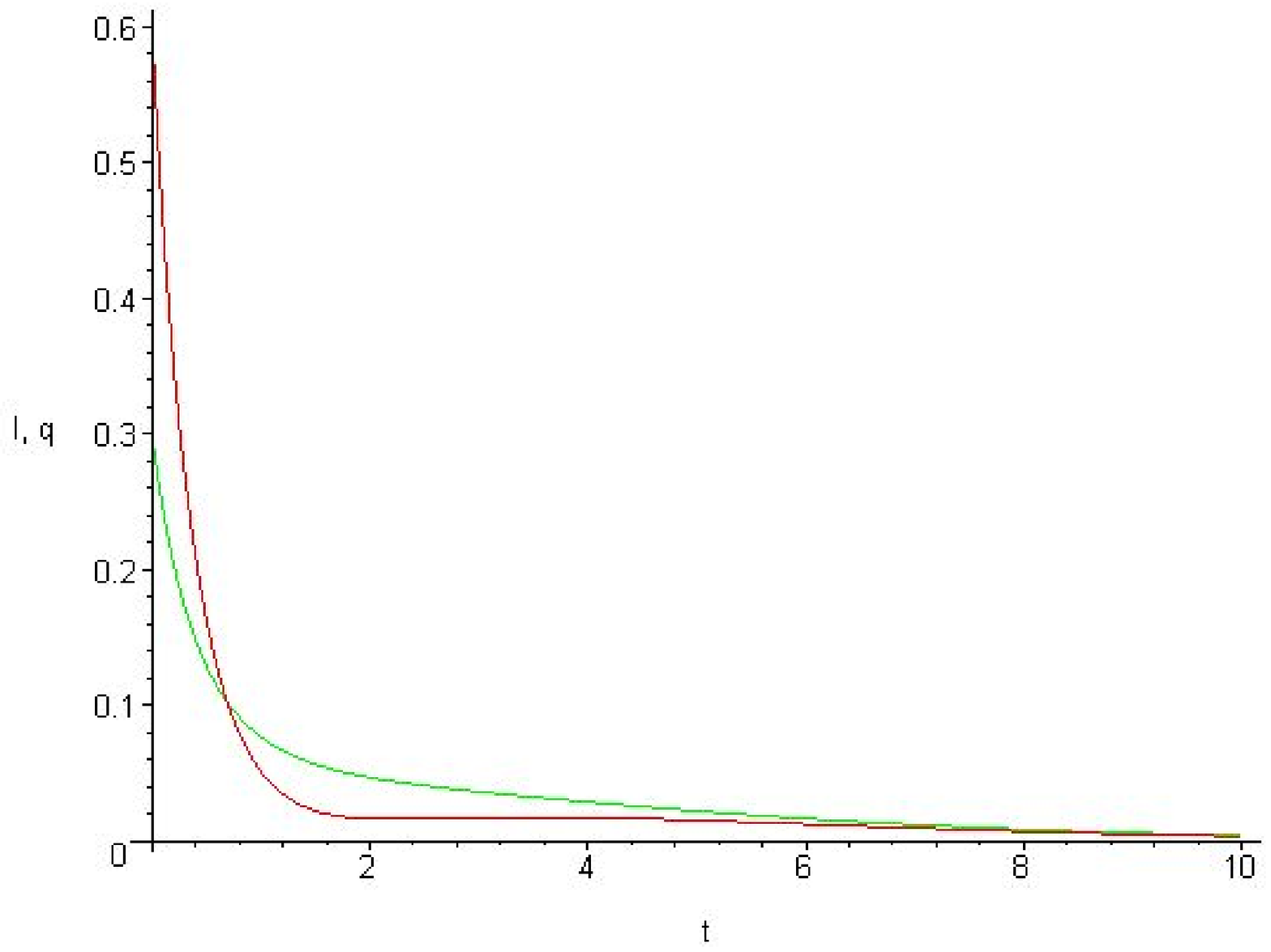}
\begin{figcaption}{swfig3}{2.5in}
A plot of the two expansion rates $l$ (upper curve for small values of $t$) and $q$
as a function of time $t$. It is seen that as the winding modes
annihilate, the expansion rates converge to each
other before converging to zero.
\end{figcaption}
\end{center}
\end{figure}
\begin{figure}
\begin{center}
\singlefig{2.5in}{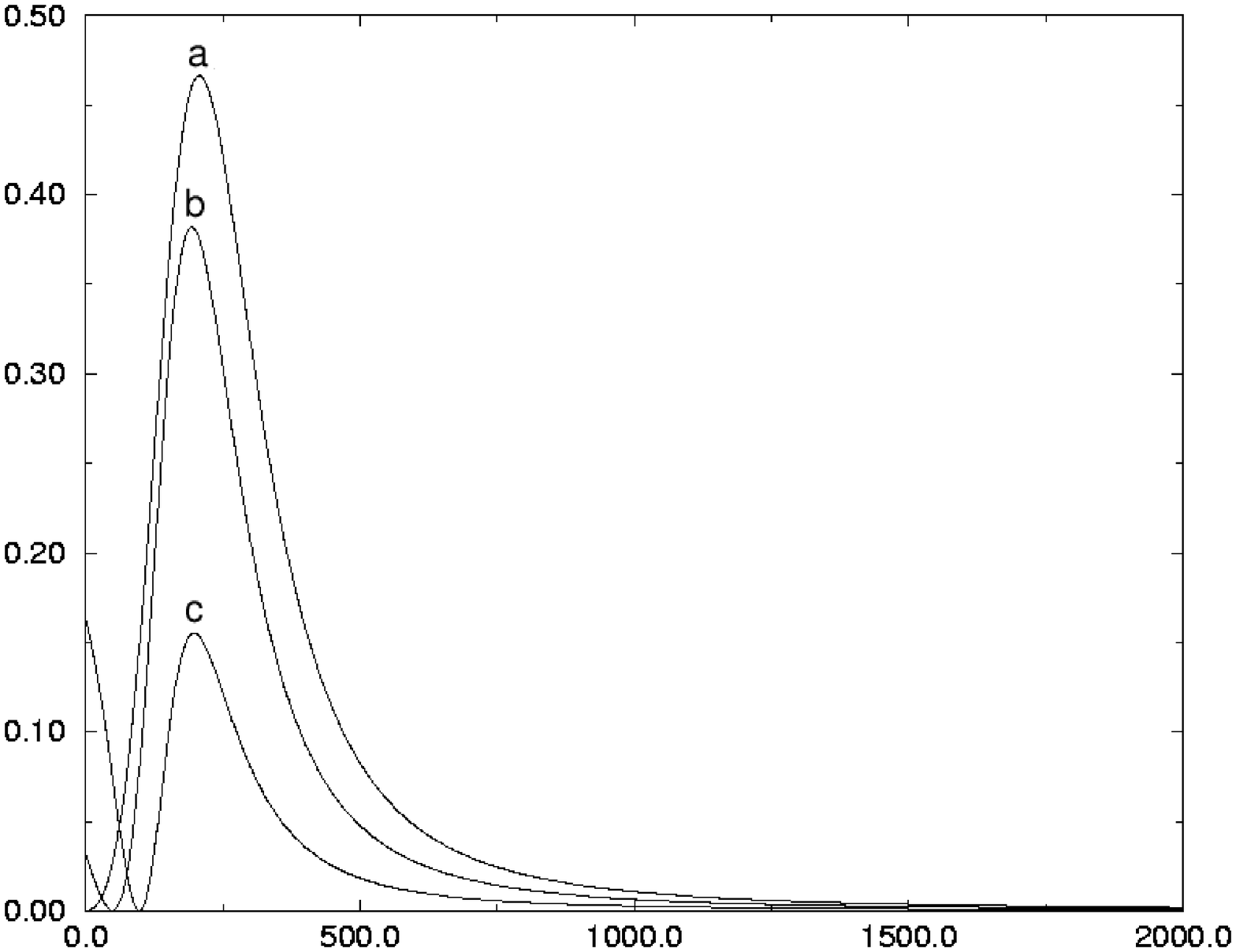}
\begin{figcaption}{swfig4}{2.5in}
A plot of the anisotropy parameter ${\cal A}$ as a function of time $t$
for a range of initial anisotropies. Curve $a$ represents no initial
anisotropy, but varying winding numbers.  Curve $b$ has an initial
anisotropy of $A=0.025$ and curve $c$ has $A=0.160$.  We see that the
anisotropy parameter reaches a maximum early on in the evolution and in all
cases tends to zero at late times.
\end{figcaption}
\end{center}
\end{figure}

However, the above
alone is not enough evidence to prove that isotropization of the
dimensions occurs.  To prove that isotropization occurs we need
to study the geometry of the background.
We can see from Figure \ref{swfig3} that the
expansion rates, $l(t)$ and $q(t)$, converge
at late times. Note, however, that at the same time the ratio of scale
factors continues to increase.  A more quantitative definition of
isotropization can
be obtained by defining the average {\em Hubble parameter}
${\bar \lambda}$ and the
{\em anisotropy parameter} ${\cal A}$ \cite{Chen:2002ks},
\bea
{\bar \lambda} \equiv \frac{1}{D} \sum_{i=1}^{D} \lambda_{i}\\
{\cal A} \equiv {1 \over D}\sum_{i=1}^{D}\frac{(\lambda_{i}-{\bar \lambda})^{2}}{{\bar
\lambda}^{2}}.
\eea
In our case we have $D=3$ and the anisotropy parameter becomes,
\beq
{\cal A}=2 \frac{(l(t)-q(t))^{2}}{(2l(t)+q(t))^{2}}.
\eeq

We find that for any amount of initial anisotropy, ${\cal A}$ approaches a maximum
value and then goes to zero, as can be seen from Figure \ref{swfig4}.
Furthermore, for the case of isotropic initial expansion
but inequivalent winding mode numbers we find that ${\cal A}$ reaches a larger
maximum but the conclusion of isotropization at later times
is unchanged.

\section{Conclusion}

We have generalized the equations of BGC to the anisotropic case
including the effects of winding state annihilation and loop production.
We address the issue of isotropization of the three large dimensions
quantitatively by introducing the anisotropy parameter ${\cal A}$.  Our
analysis indicates that for an arbitrary amount of initial
anisotropy, the anisotropy will reach a maximum early in the
evolution and then approach zero at later times. Thus, we have
explained how isotropy arises as a natural consequence of BGC.

\begin{acknowledgments}

RB was supported in part by the U.S. Department of Energy under
Contract DE-FG02-91ER40688, TASK A.
SW was supported in part by the
NASA Graduate Student Research Program.
RB also wishes to thank the CERN Theory Division and the
Institut d'Astrophysique de Paris for hospitality and
financial support during the time the work on this project was
carried out.

\end{acknowledgments}


\end{document}